# Knowledge discovery via multidimensional science maps: the case of the Species Problem

**Sándor Soós**

Dept. of Science Policy and Scientometrics, Library and Information Centre of the Hungarian Academy of Sciences (MTA)

**Abstract**

Science mapping (SM), the study of the organization and development of science and technology, is a rapidly developing field within information science. The volume of available data allows this methodology to empirically address such issues as the historical development of topics, discourses, fields or of the entire science system. Based on the pool of related methods, we are proposing an integration of various maps to obtain a novel kind of science map we call multidimensional. The basic idea behind is to combine the most informative relations available from various maps based on different bibliometric indicators, in order to produce a rich structrue for the study of knowledge dynamics, with special emphasis on causal–historical connections. As a proof of concept, we deploy the proposed framework in an extensive case study on a historical topic from the life sciences, namely, the debate on the species concept in biosystematics.

**Keywords**

Science mapping, multiplex networks, knowledge discovery, species problem, history of biology

**Introduction**

Science mapping (SM), the study of the organization and development of science and technology, is a rapidly developing field within information science. With the contemporary supply of extensive bibliographic databases, the volume of available data allows this methodology to address such issues as the historical development of topics, discourses, fields or the entire science system. The main aim of this paper is to elaborate on existing mapping methods, rooted in network science, via a concept we call *multidimensional science map*. An equally important aim is to start providing a proof of this concept via an extensive case study on a historical topic from the life sciences.

Network methodologies for uncovering the organization of scientific aggregates are best arranged along two dimensions at the operational level: (1) the type of bibliometric indicator(s) used, and (2) the model built, based on the selected indicator(s). Main categories of models and their corresponding applications are the following.

- Methods of reference pattern mapping

The most paradigmatic methodology in the bibliometric mapping of the structure of scientific fields is based on the indicator set provided by the references of papers. The main assumption behind this is that references jointly constitute the intellectual background or the „knowledge base" of papers: therefore, analyzing the aggregated reference set of a corpus representing a field at any given time slice uncovers its cognitive structure. References provide multiple bibliometric indicators for analysis, including
- cited documents (D),
- cited authors (A),
- cited sources (typically journals) (S),

each conveying a different aspect of the field structure (see below), and posing different requirements on, or challenges for computing power. Methods utilizing the above indicators fall into two major categories.

*Bibliometric coupling* (BC). Source documents representing a field are clustered based on their degree of sharing the same references (in terms of documents, authors or sources). Various measures of similarity and clustering techniques are used.

*Co-citation analysis* (CC). References in source documents are clustered based on their frequency of being co-cited by the source document set (in terms of full references, included authors or sources). Again, various measures of similarity and clustering techniques are used.

The two basic techniques are the converse of each other: in BC, source documents are grouped via references; in CC references are grouped via citing source documents. Prototypic approaches addressing the organization of science using reference-based mapping,are the following.

*Intellectual structure of fields.* A rather traditional approach, author-co citation analysis (ACA) is often used to detect and visualize the cognitive structure of research fields. ACA is the combination of (A) and (CC), as it takes cited authors as the unit of analysis, and yields author clusters based on their "co-citedness". Clusters are conceptualized as research communities concentrated around a specific research topic, thereby mirroring the thematic composition of the underlying field.

*Disciplinary organization of science.* Variants of source co-citation analysis (SCA) are used in large-scale approaches addressing the global structure of science. For constructing a "global science map", several researchers used SCA. In models of Moya-Anegon *et al.* (2004), specialties of science represented by Subject Categories in the ISI databases are subjected to analysis. Subject Categories are source indicators, as they are introduced to categorize journals in the database. Using source documents, and substituting cited sources for the corresponding Category, the proximity of Subject Categories is calculated measuring the degree of their co-citedness throughout the whole corpus. As a result, a proximity network of Categories (specialties) is obtained, which can represent the global structure of science.

The method described above is thus a combination of (S) and (CC). A somewhat different approach has been introduced by Leydesdorff and Rafols (2009), whereby Subject Categories are related upon their citation patterns, namely, by their degree of co-citing the same Subject Categories, resulting also in a proximity network. Consequently, while the maps of Moya-Anegon *et al.* belong to the class of co-citation analyses, the latter approach is an example of bibliographic coupling on (aggregated) sources, i.e. a combination of (S) and (BC).

*Global map of scientific paradigms.* The most detailed picturing of the scientific landscape  to date has been achieved by the "paradigm mapping method" (Boyack et al 2005, Boyack 2009). A paradigm in this setting is operationalized as a frequently co-cited group of references, reflecting a cohesive topic or specific subject of research. The global paradigm map of (Boyack et al. 2005) has been generated by processing the content of the Scopus database: the full references of source documents were subjected to co-citation analysis, and clustered based on the resulting proximity matrix. The procedure yielded something that may be considered the global map of scientific paradigms at a given time slice, unraveling a cluster structure at an extremely high level of granularity. Since the method relies on full references, it qualifies as an instance of document co-citation analysis (DCA), hence combining (D) and (CC). It should be noted that, due to the outstanding amount of documents and references, the method requires considerable computing power.

- Citation-flow mapping

By utilizing the nature of scholarly citation, corresponding indicators naturally enable science mapping to empirically address dynamic or historic aspects of science. To detect and visualize the flow of information, the spread and transformation of ideas, or the development of conceptual systems, citation-flow mapping is utilized, sometimes called „algorithmic historiography" (Garfield et al. 2002). The method implies the construction of a citation network of papers either over a given timescale or about a given topic. This network is conceptualized as representing the patterns of information flow. The network, in this case, is static in the graph-theoretic sense, but represents longitudinal, i.e. dynamic content, mapping a process along its time dimension. A genuine implementation of the concept can be found in *HistCite*, a software providing citation flow analysis based on the ISI databases.

- Author-based mapping

Another common bibliometric indicator for science mapping is authorship, or, rather, co-authorship. Many studies have attempted to reveal the composition and development of research fields by analyzing the author-network encoded in the corresponding publication corpus. The collaboration of authors resulting in joint publications is conceived as a shared research interest, upon which the „visible colleges" of a field or a discipline can be identified. Technically, the analysis of co-authorship patterns proceeds by first extracting the network of authors from a bibliographic dataset, where ties stand for two actors co-authoring at least one paper. This network is then subjected to community detection methods, and decomposed into coherent author clusters, that is, into scientific communities. Though this approach is, at face value, just an application of social network analysis, and so targeted at the social organization of science, the factors behind group formation, such as working on close topics, make it capable to grasp cognitive organization as well.

Co-author networks are often studied from within the network science perspective, irrespective of their use for science mapping purposes. Various generalizations have been made on the structure and dynamics of such networks, drawn from assigning them to the class of scale-free networks. For instance, the growth of co-author networks by „preferential attachment", a process responsible for many scale-free structures, is also a well-known claim (Barabasi et al. 2002). Other studies more directly in the SM domain have investigated general co-authorship patterns and -dynamics in relation to the evolution of research fields (Bettencourt et al. 2009). The report below will heavily utilize the results of this latter approach.

- Mapping conceptual structures

A different, and frequently utilized set of bibliometric indicators is constituted by the textual descriptors of documents. Descriptors in this category include keywords associated with documents, title words, or the characteristic words obtained by text mining from either the abstract or the full text of papers. As can be seen from the list, the methodology may involve natural language processing and text mining procedures, which makes this approach relatively expensive compared to the utilization of directly accessible metadata types. For the sake of simplicity, we describe the methodology below using the case of author keyword analysis. Author keywords are concepts chosen by the author to jointly convey the content of the respective paper, being readily available in many scholarly databases among the metadata of documents. Therefore, processing author keywords does not require text mining or other linguistic pre-processing.

Since keywords are meant to provide immediate access to the content of papers, their association patterns in large-scale document sets are considered as (1) the most directly interpretable and (2) the most fine-grained mapping of the cognitive structure of the underlying field. The method is referred to as co-word analysis: first, a pairwise association of keywords is measured in a document set, based on the frequency of their co-occurring in documents. Next, this association matrix is decomposed, either by direct clustering or (conceived as a proximity network of concepts) by community detection methods yielding groups of closely related words. These groups are then interpreted as thematic clusters comprising the field under study.

Co-word analysis, the alternative method for building a representation of the cognitive structure of science, is often contrasted with co-citation analysis as being suited to somewhat different tasks of science mapping. The main argument is based on the recognition that references encode the past, or background of a paper, while keywords are „of the same age" as the source document itself. Hence, co-word analysis is argued to be more capable of grasping ongoing trends or emergent topics than co-citation patterns, which may not react to rapid changes or to the appearance of genuinely new directions (cf. Chen 2003). Sensitive as it is, the co-word approach has also been challenged by theoreticians such as Leydesdorff (1997) who pointed out that, among other things, the association of words without a sufficient information about the embedding context leads to uncertain interpretations and risky semantics, violating the validity of evaluating these maps.

To achieve higher levels of accuracy and expressive power, the basic methods summarized above are also often combined, resulting in various hybrid methods. (cf. Janssens 2008).

**Methods: multidimensional science maps**

Based on the pool of related methods described above, we are proposing an integration of various maps to obtain a novel kind of science map we call *multidimensional*. The basic idea behind this proposal is to combine the most informative relations available from multiple maps based on different bibliometric indicators, in order to produce a rich structrue for the study of knowledge dynamics, with special emphasis on causal–historical connections. In particular, given any publication record *P*, our model consists of the set-theoretic union of three graphs extracted from *P*:

1) *Author-citation network induced by P.* The directed and weighted graph representing citation relations in *P* among authors within *P*.

2) *Keyword-citation network induced by P.* This rather unusual type covers the citation relations among key concepts within the corpus, based on the citation network of documents in *P*. In other words, this type of map is to reveal the descendancy of concepts and the development of the conceptual system based on actual knowledge flow. The network is a directed and weighted graph.

3) *Author-keyword network induced by P.* This map type differs, in terms of network theory, from both types 1–2 in that it is a so-called bipartite graph: it relates two different indicator set, that of authors and keywords. Practically, this bipartite graph creates a mapping betwenn the previous two network types, as it is to be induced by the author/keyword set within *P*.

To put it differently, our proposed model links or „matches" the knowledge flow among authors and concepts in a single representation via connecting the respective two graphs by a third one, that is a coupling of authors and concepts. We argue that integrating this three bibliometric aspects of scientific discourses, or three traditional types of science maps has various benefits in the study of knowledge dynamics:

- *Semantically informative structure.* Traditional citation networks are, in most cases, difficult to interpret even if a tractable structure is detected in the graph. The primary reason is that widely used author citation networks speak of „formal historiography" in terms of (proper) names, therefore, interpreting the history requires additional sources of information on related concepts, ideas etc. In the network studied below, parallel citation networks induced by authors and concepts are linked together, ensuring a semantics for the analyst to author descendencies and interrelations identified in the graph, as a key to interpret underlying traditions.

- *Filling the gaps of missing links/data.* An inherent feature of „unidimensional" networks, especially in the case of keyword nets, that the underlying dataset is a partial one: in historical publication records, for example, older publications usually miss associated keywords or other content descriptors, typically due to a database/indexer effect. When, however, connecting author and keyword nets, the complementer relation, that is, citations between authors, may fill the gap of missing citation links between concepts. Consequently, (historically) related sets of authors and terms may reveal themselves as cohesive groups to, e.g. community detection methods (see below), even in the absence of explicit relations on either side.

- *Historical (causal) relations instead of co-occurence.* A feature of high importance associated with the multidimensional maps is that is is constructed out of citation relations, that is, causal links in each dimension. Traditional concept maps are induced upon the co-occurence of keywords in documents, which is a useful indicator of topics, but stillan associative approach missing actual causal links or descendancy relations conveying the paths of knowledge flow. In our map, keywords a related through citation relations, allowing the analyst to directly track the evolution of the underlying conceptual system.

In the rest of the paper, we present a case study of the application of this technique, the main goal of which is to demonstrate, through uncovering the historical structure of a complex scientific discourse, the benefits of the approach. Our methodology in implementing the multimap proposal consisted of the following steps:

1) In the first step, based on a large-scale corpus collected in relation to the topic (see below), we obtained the three constituent map of the publication record, that is, the author–citation network, the keyword–citation network, and the author-keyword graph.

2) In the next step, the three graphs have been unified along common nodes (set-theoretically), resulting in the final, multirelational network.

3) We have filtered and normalized the raw multinetwork in a variety of ways, to adjust for the differences between the traditional graphs. Most importantly, edge weights have been normalized to range from 0 to 1 in each constituent graph, individually, since e.g. the frequent relations characteristic between keywords would have suppressed the much weaker associations in the author–keyword graph.

4) As the definitive step, in order to reveal the structure of the discourse, we have identified research traditions as subdiscouses in the network as cohesive subgraphs via a community detection algorithm based on modularity maximization.

The method of community detection applied here is, in principle, the result of integrating two approaches aiding at community detection in complex networks. The algorithm attempts to identify communities mostly based on the topology of the underlying graph, so that the resulting groups can be characterized as maximizing within-community connections, while minimizing inter-community connections:

(1) the Walktrap Community Findig (WCF) algorithm attempts to find dense subgraphs within a network by random walks (Pons & Latapy, 2005). The underlying idea for this algorithm is that short random walks with the probabilities determined by the edge weights are likely to circumscribe a community in the sense of being a set of densely and strongly connected nodes. The WCF algorithm works in an agglomerative fashion, starting with the strongest communities and merging the closest ones in consecutive steps until the whole network is reconstructed.

(2) The iterative procedure (1) is repeated until an optimal community structure is obtained. A now-standard method for optimization is the application of the network measure called modularity (Newman, 2006):

$$Q = \frac{1}{2m} \sum_{i,j} A_{ij} - \frac{k_i k_j}{2m} \delta(c_i, c_j),$$

where $m$ is the number of edges, $A_{i,j}$ is the corresponding element (weight) of the similarity matrix, $k_i$ and $k_j$ are the degrees of the corresponding nodes, $c_i$ and $c_j$ are the community indices the two node belongs to, respectively. $\delta(c_i, c_j)$ is a function that equals to 1 where both nodes are of the same community *(c_i = c_j)*, and 0 otherwise. Informally speaking, the function measures how "modular" a given network is under a certain partition of its nodes (community structure), i.e. how separated the different node types (communities) are from each other. Using this measure as the object function to be maximized, that is, by $Q \rightarrow \max$, the algorithm identifies the optimal (most modular) partition of the network (without putting artificial constraints on CD, such as similarity thresholds).

## Materials: a corpus on the Species Problem

In order to test and demonstrate the capacity of the proposed method, we applied it in an attempt to reconstruct the historical development of a rather complex discourse in biology, usually referred to as the *Species Problem*. The Species Problem can be briefly described as a historical debate on what biological species are, and as the related quest for the appropriate definition of species, or species concept for biology. With a long prehistory, dated as back as to Aristotle and Plato, including Darwin's paradigm-shifting work on the nature of species in the XIX. century (milestone #1), the debate expanded in the early XX. century, mainly due to the rediscovery of Darwin's work, and having it integrated with the early (Mendelian) genetics of the era. The new paradigm has been called the *Evolutionary Synthesis* (milestone #2). Since the Synthesis, a plethora of theories has emerged on species, resulting in a variety of competing species concepts. According to a comprehensive review of Mayden (1997), no less than 22 species concept (definitions) exhibit themselves in the contemporary literature of the subject.

Given its complexities, the Species Problem was an ideal candidate for a bibliometric analysis of —inter-, or multidisciplinary—knowledge diffusion with the proposed methodology:

(1) The roots of the discourse are centuries-old, while there are several contemporary directions of the debate (and of research) as well (cf. Hull 1988, Ereshefsky 1992).

(2) During its modern history (in the XX. century), many schools of biosystematics contributed to, and competed over the problem, involving—from a data-mining perspective— different topics: theoretical papers as well as empirical ones, the latter focusing on particular subjects of taxonomy (description of taxa). It was of outstanding interest whether the enhanced overlay toolkit was capable of identifying these knowledge transfer induced by the interaction of these schools.

(3) A nonstandard feature of the Species Problem is its complexity in terms of the contributing scholarly fields, or even disciplines. For example, a proper interaction of evolutionary systematics, on one side, and the philosophy of science (of biology), on the other side, had a significant effect on the present state of the debate. It is a good challenge for the proposed method of mapping science dynamics to capture the associated degree of knowledge diffusion being often discussed by historians of science.

To cover a representative corpus of the modern history of the discourse, bibliographic data were harvested from three databases of the Web of Science, namely, the SCI, the SSCI and the A&HCI (*Science Citation Index*, *Social Science Citation Index* and the *Arts&Humanities Citation Index*, respectively). Also in a attempt to avoid the potential exclusion of relevant works from the corpus, data retrieval was based on a topic-related query, that did not put any constraints on the set of fields, journals, authors etc. entering the sample. The query was defined to include all records related (topicwise) to any of the following terms: „species problem", „species definition", „species concept".

The resulting initial or „core" corpus included N=1605 documents for the period 1975–2012. In an attempt to gain a comprehensive historical coverage on the topic, we have iteratively extended our initial set via an in-depth analysis of aggregated references included in the core set. In the first step of the process, references from the initial corpus were processed and obtained (as source documents) from the WoS databases. This additional publication record was then added to the pool of already collected papers. We repeated this method in further iterations, until reaching a collection being fairly „closed" under the citing relation, that is, a collection that contained all the—topic-relevant—papers referred in the discourse. To assure such a convergence, references were filtered by a threshold imposed on their frequency: papers cited above this threshold were, in each round, considered *relevant* for the topic. The threshold value was increased (non-linearly) for each iteration, based on the assumption that the farther we get, along a series of references, from the core set of papers (in terms of corpus generations), the less related references will be to the topic. Interestingly, with this setting, the procedure converged in the third generation of papers, indicating that almost all relevant references were present after two iterations. Finally, for the discourse of the species problem, we arrived at a final record of approximately 5700 papers (the main statistics of the procedure are summarized in Table 1.)

**Table 1.** *Statistics of iterative corpus collection on the Species Problem based on WoS databases*

| Iteration | No. of source documents | No. of references | No. of unique references | Threshold value | No. of relevant references (retrieveable) |
|---|---|---|---|---|---|
| Initial corpus | 1605 | 93 943 | 50 668 | 3 | 3223 |
| 2. generation | 3223 | 155 742 | 62 574 | 10 | 851 |
| 3. generation | 851 | 14 991 | 5305 | 10 | 2 |
| **Total** | **5679** | | | | |

To prepare the knowledge discovery along this large-scale longitudinal bibliography, we have organized the final corpus into a citation network. The large directed graph obtained from the document set consisted of all papers included in the full corpus as its nodes; edges represented the (direct) citing relation between any two documents. The rest of the paper reports the comparative (longitudinal) analysis of this network via analyzing it in the form of a multidimensional map.

## Results and discussion

The community detection on the combined author–keyword citation network resulted in 5 major coherent groups, that is, five major discourses could be identified within the history of the problem. These discourses — modularity classes — are presented below in two, complementry ways: for each identified module the the graph is presented (visualized) in a reduced form, omitting less connected nodes for better readability. At the same time, as to the quantititive version, the most important nodes (authors/keywords) based on their PageRank centrality are plotted in the form of a barchart, characterizing the author group and the conceptual system of the module.

### 1. *The phylogenetic and cladistic theory of the species category.*

The most extensive discussion, accounting for the largest module in the graph, may clearly be interpreted as the theoretical debate focusing on a species category defined in terms of phylogenetic criteria and theory. By the reduced graph (Fig 6), two qualified species concepts show itself as organizing the discourse: the *Phylogenetic Species Concept*, and the *Genetic Species Concept*. Even more telling is the structure of the subgraph, as evidenced by both the visualization and the centrality-ranking of authors/concepts depicted in Fig 1. The upper part of of the graph (*Ereshefsky, M*, *definition*, *clade*, etc.) mirrors the contribution of philosophers of science and theoreticians of biology to concept formation: the concentration of these approaches is rather striking in the full network of this module (Fig. 7, framed area), whereby most influential „philosophers" of the problem are present  (*Ghiselin, Hull, Wiley, Sober, Mishler, DeQuiroz, Platnick, Cracraft* etc.), along with a set of thematically related key concepts on the ontology of species (*individual, class, definition, ostensive definition, name*). This group is connected, through a set of central concepts (including concepts from experimental science, such as *mithocondrial DNA*, *DNA barcoding*) to an extended group of approaches addressing species within experimental/molecular biology. The so-called *genetic species concept* is positioned in this context, while the *phylogenetic species concept*, as such, is positioned in the neighborhood of theoreticians. This configuration of the network corroborates, on one hand, (1) the substantial — interdisciplinary — interaction between the philosophy of science and species systematics. The famous *individuality thesis*, stating that species are ontological individuals instead of classes, is, indeed, seems to penetrate the discussion on the species category, serving as the philosophical background for the *phylogenetic* and—as a highly related definition— the cladistic concept. This interdisciplinarity is also made apparent by the centrality ranking of network members: the high end of the distribution shows *Hull, DL*, the philosopher co-inventor of the individuality thesis along with *mithocondrial DNA* as the third and second most central actor in the net, respectively. It is also of great interest that the tradition of theorizing on the species category, the majority of „philosophers" of the issue, show up almost exclusively in this subdiscourse, that is, in relation to the phylogenetic conception. On the other hand (2), a further important historical connection emerges from this module, between a theoretical and an experimental tradition. Based on the network structure outlined above, it can be hypothesized that the genetic species concept is a a descendant of the phylogenetic species concept, the former being an operationalized or, at least, more applicable version of the latter in the context of experimental, namely molecular, biology.

## 2. *Research on phylogenetic inference*

The next module in the list, in terms of graph size, is a well-interpretable and highly coherent research tradition overlapping with the quest for a valid species category. The reduced graph (Fig. 8) reveals the discourse on *phylogenetic inference*, that is, the methodology on experimentally inferring and reconstructing phylogenies of/among taxa, including species. *Phyologenetic inference* is both a methodological and experimental subject within evolutionary biology, as is clearly reflected in the set of constitutent concepts. The structure of the module is indicative of both its relation to the species problem, and of its coherence: through central concepts (*phylogeny, species delimitation, molecular systematics, molecular phylogeny*) two cohesive groups are connected to each other: the set of authors interacting on this subject, and the related conceptual system as an apparent description of the methodological issues involved. The interface of this tradition with the species problem is the valid procedure of experimentally delimiting species (by phylogenetic reconstruction): most interestingly, this experimental methodology applies multiple theoretical species concepts (as evidenced by the nodes *morphological species concept*, *biological species, phylogenetic species*) for the purposes of operationalization. This methodological character is also evidenced by the Page Rank centrality ranking (Fig 2), whereby *parsimony*, as the main axiom or object function of the inference method is shown as far the most central concept, along with the author *Felsenstein*, known for the firs phylogenetic inference software package. Even the long tail of the centrality distribution almost uniformly covers mathematical and experimental methods (*weighted/unweighted least squares, maximum likelihood, Bayesian estimation* etc.). Though not apparent either on the reduced layout or in the ranking plot, by a relatively weak link, Ernst Mayr, the classic figure of the species problem originally proposing the biological species concept, is also classified together with this module. The connection is established through the concept *natural system* (present in the reduced graph), a Darwinian principle rediscovered by Mayr for systematic biology, and—apparently—entertained by this tradition as the primary criterion for selecting among alternative inference methods.

## 3. *Speciation and the BSC tradition*

The context one would expect Mayr to appear within would be the next most significant subdiscoruse, which altogether can be referred to as the tradition induced by the Biological Species Concept (BSC). The subgraph, again, mirrors a highly cohesive group (Fig. 10): a densely connected set of concepts is being related to a set of interacting authors through, basically, three central terms: *speciation, reproductive isolation* and *hybridization.* Almost all concepts are clearly related to an aspect of the debate on the biological, or interbreeding-based definition of the species category: e.g. *sexual isolation, hybrid inviability, hybrid sterility, gene flow, ring species.* The same phenomenon is being shown via the centrality ranking plot (Fig. 4): each constituent in the list of terms is related either to a classic featrure of the BSC, or to the original arguments in support of the conception. The most central term, however, is *speciation,* the mechanism of species formation which, with the Biological Concept, took a definitive role as a phenomenon that any theoretically sound species concept should explain. In sum, a natural interpretation of this module is that it is organized by the debate on speciation as framed by the BSC, with all its empirical difficulties caused by the primary criterion of reproductive compatibility/isolation. Having both field science (*vocalizations, sunflower, allozymes*) and theory engaged in the same tradition, the graph also shows the related philosophical influence on the debate: the top part of the reduced network contains *essentialism, natural kinds, levels of selection* in the neighborhood of the *biological species concept*, with related theoreticians (*Hey, Wilkins*). Though the biological conception is often communicated by historians/philosophers of science as the „death of essentialism" (whereby species taxa are no longer natural kinds), these ontological arguments are usually linked to the whole modern history of the species debate: the present result, however,  bounds the context of (explicit and terminologically detectable) anti-essentialism more closely to the BSC tradition, which is an additional piece in the historical mosaic of the species problem.

4. *Challenges for species concepts on the part of microbiology*

The next two modules, though significant in size individually, are best described in a parallel manner, the reason being both represent the same type of contribution to the species problem. As wittnessed by historians of biology, theoretically grounded and general species concepts have often been challenged from within different fields of application or the practice of systematics. Especially resistant to definitional approaches is the field of microbiology, as for example in the realm of microorganisms—mostly lacking sexual reproduction—the biological species concept, as such, can hardly work. The two subdiscourses in question cover a related research subject in microbiology, respectively, each of which poses a challenge for theoretical definitions of the species category. Both modules, therefore, convey the reception of the theoretical debate in experimental science. The more extensive (Fig. 9) is held together by the central concepts *recombination, evolution, species concept, lateral gene transfer*, which is also confirmed by the centrality ranking (Fig. 3), complementing the list with *linkage disequilibrium, bacteria.* The microbiological character of this discourse is reflected in that most constituent terms (the author interaction part aside) are names of microbial taxa. This structure is a good characterization of a quest for a microbial species concept based on phenomena in among microorganisms (mostly bacteria) that are comparable to theoretical species critera (as e.g. „recombination through lateral gene transfer"). Even more specific is the other module categorized under these approaches, concerned with a certain taxonomic group called *Diatoms* (Fig. 11). Diatoms are a type of phytoplankton or algae, that is also hard to reconcile with existing species definitions. The corresponding subgraph exhibits a set methods from cell and molecular biology aimed at the task of species delimitation.

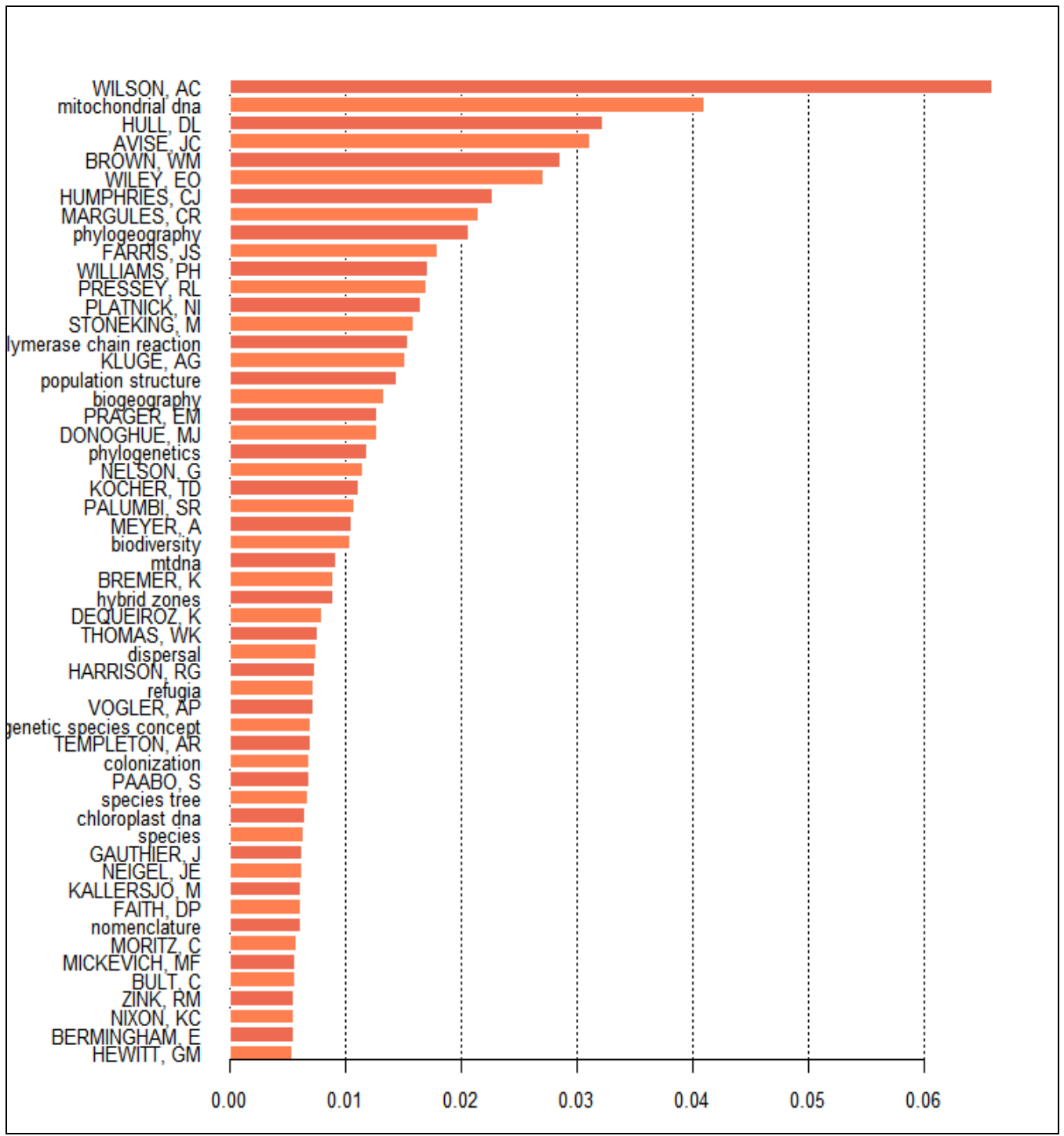

Fig 1. The phylogenetics module

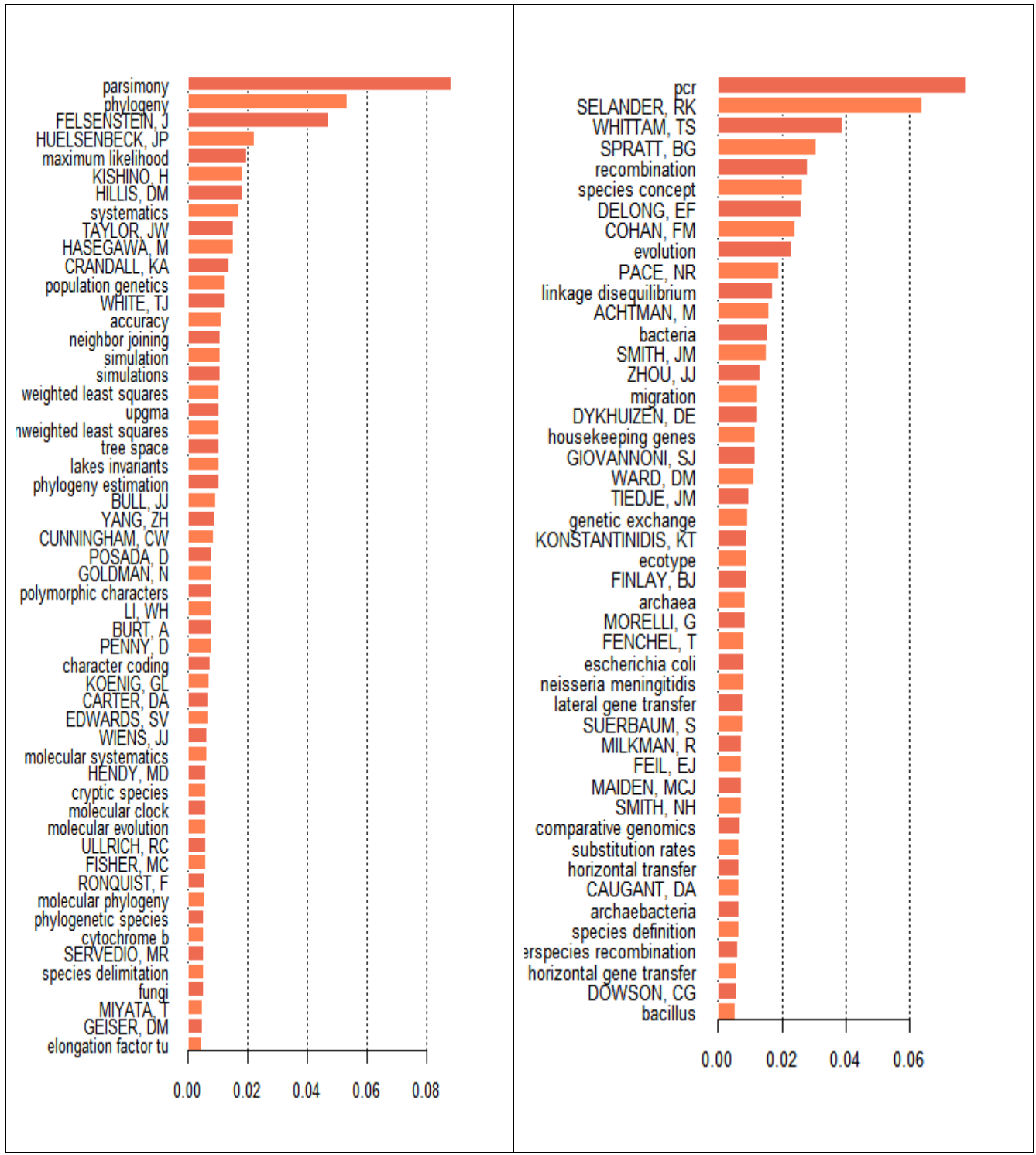

Fig 2-3. The phylogenetic inference and recombination module

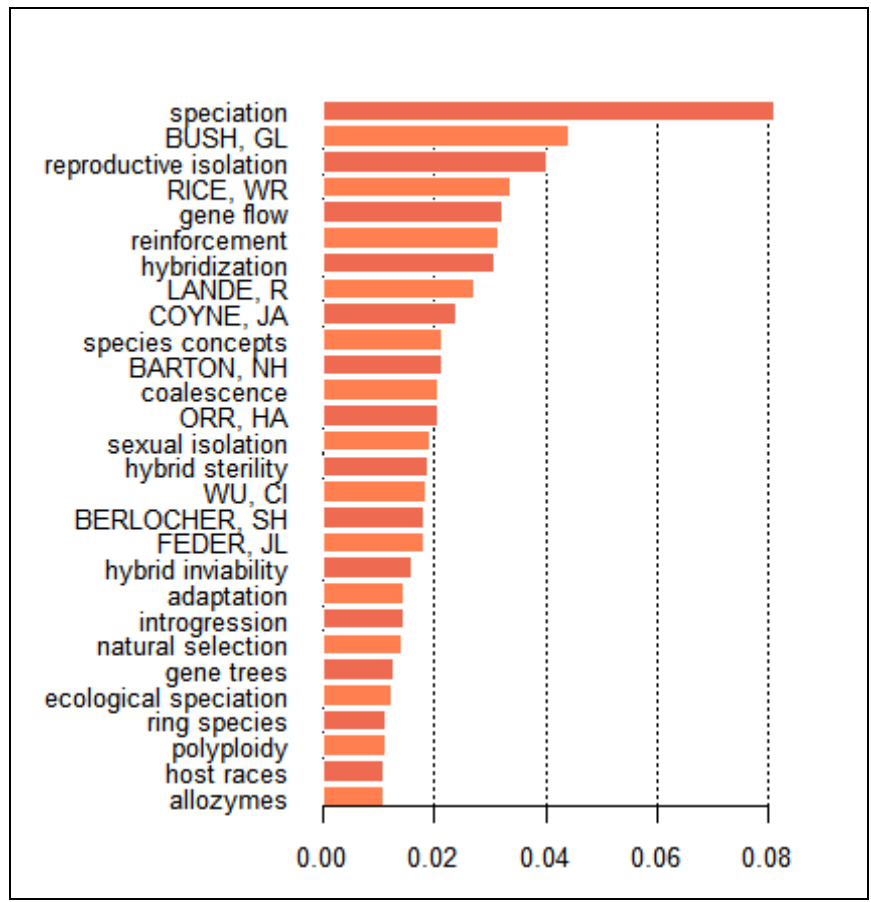

Fig 4. The speciation/BSC module

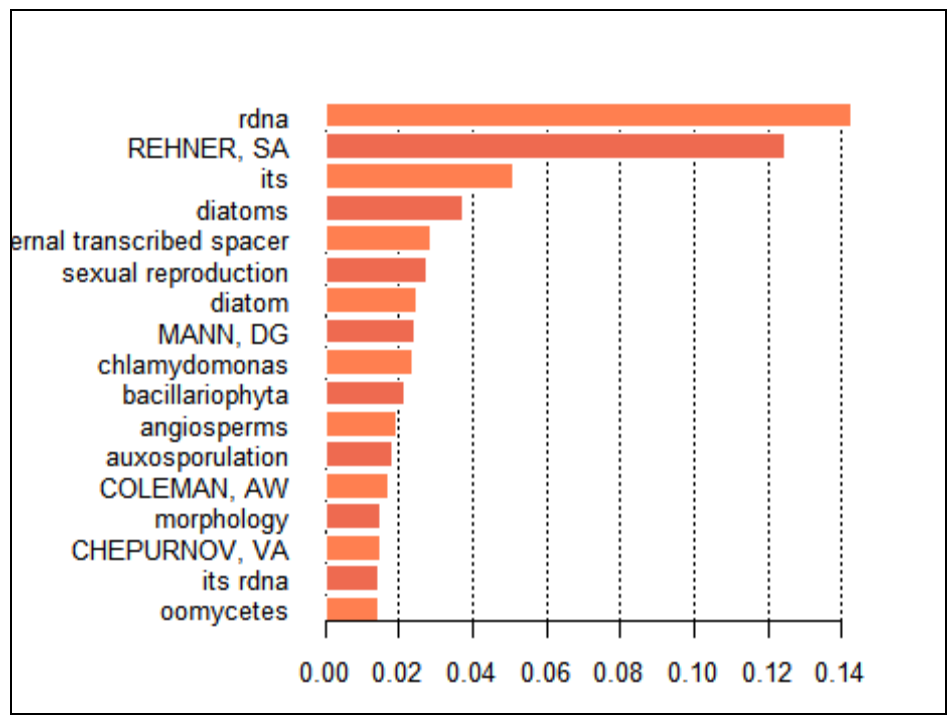

Fig 5. The diatoms module

Fig 6. The phylogenetics subgraph

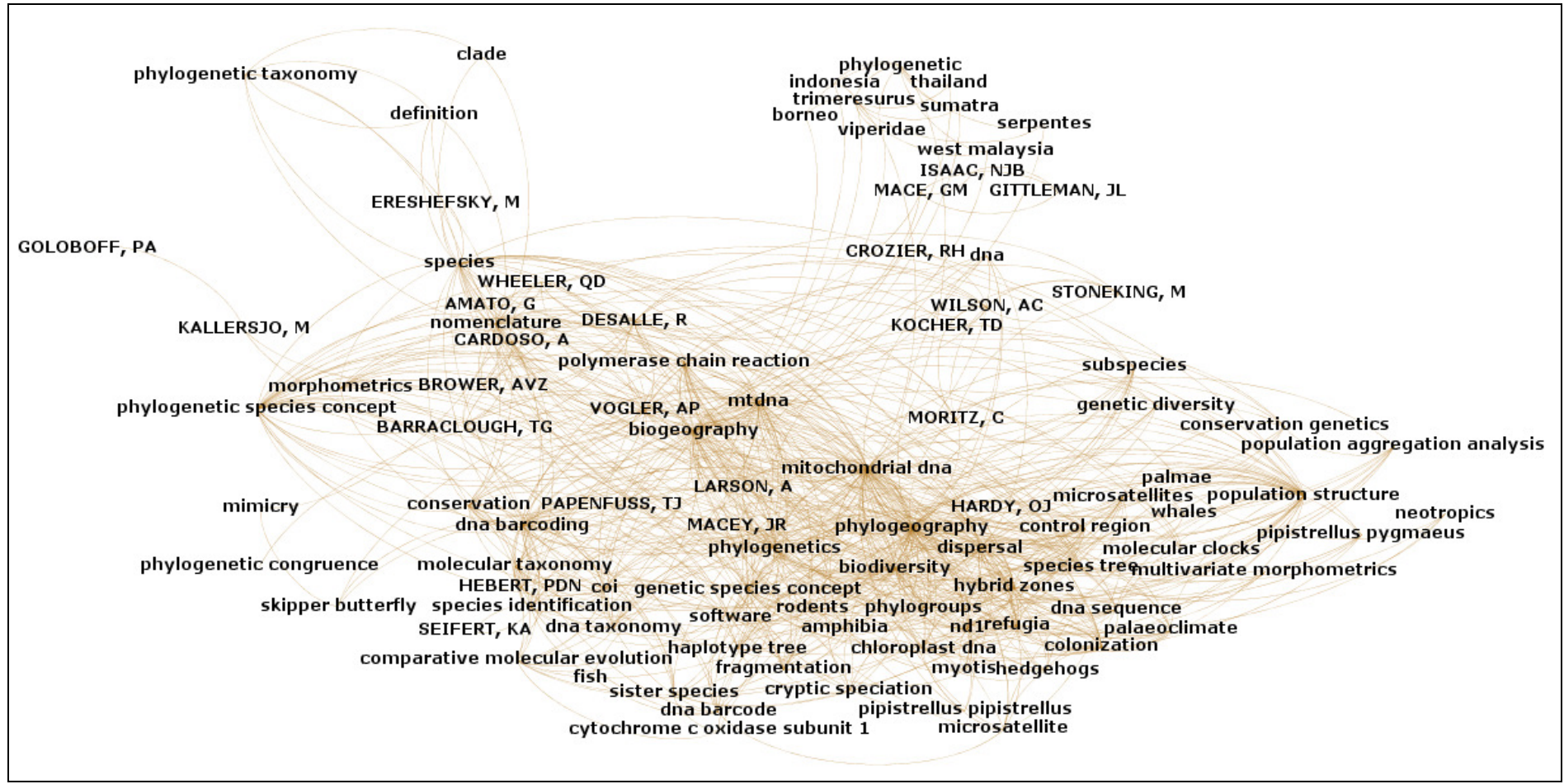

Fig 7. The phylogenetics subgraph, full version

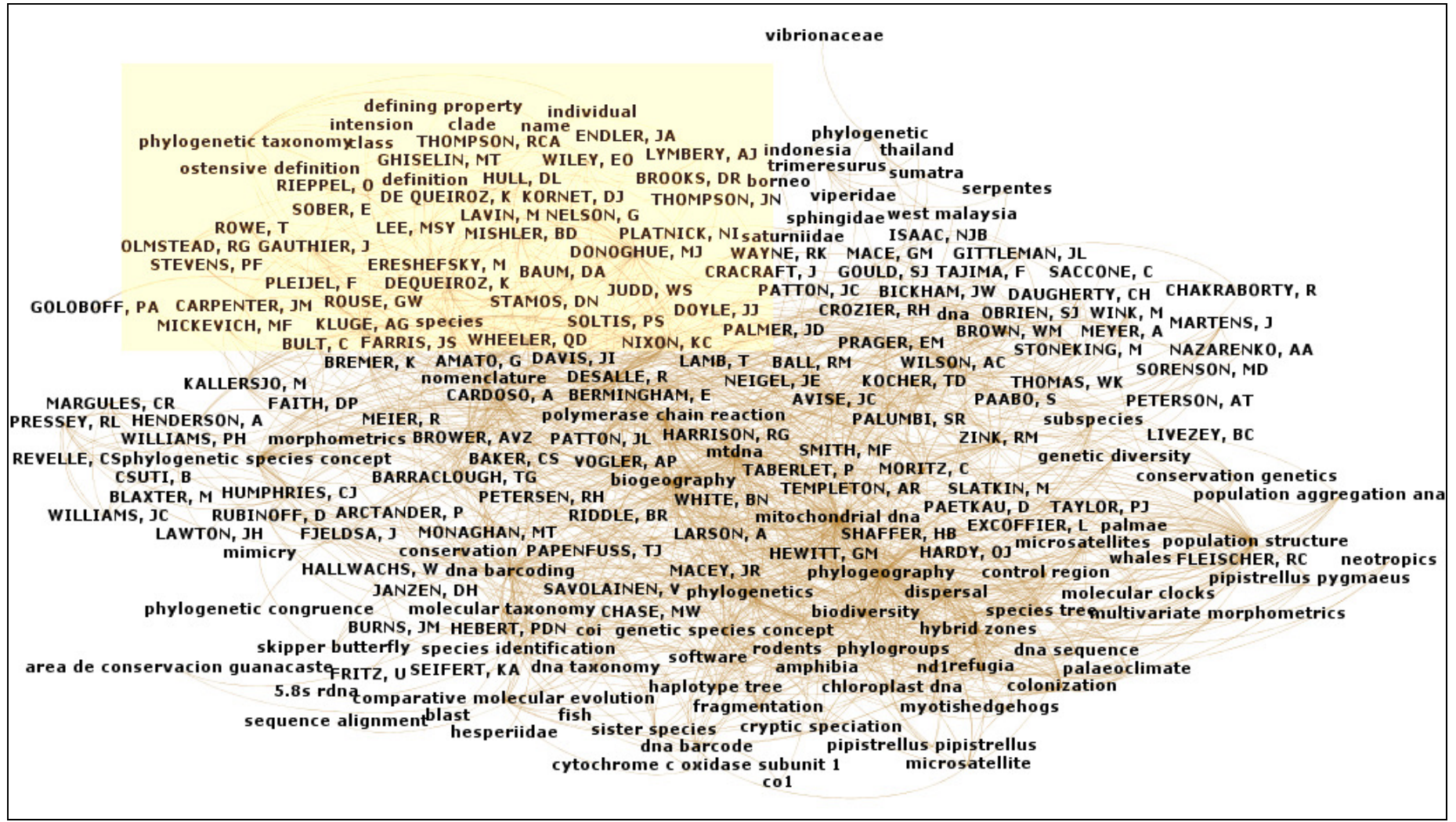

Fig 8. The phylogenetic inference subgraph

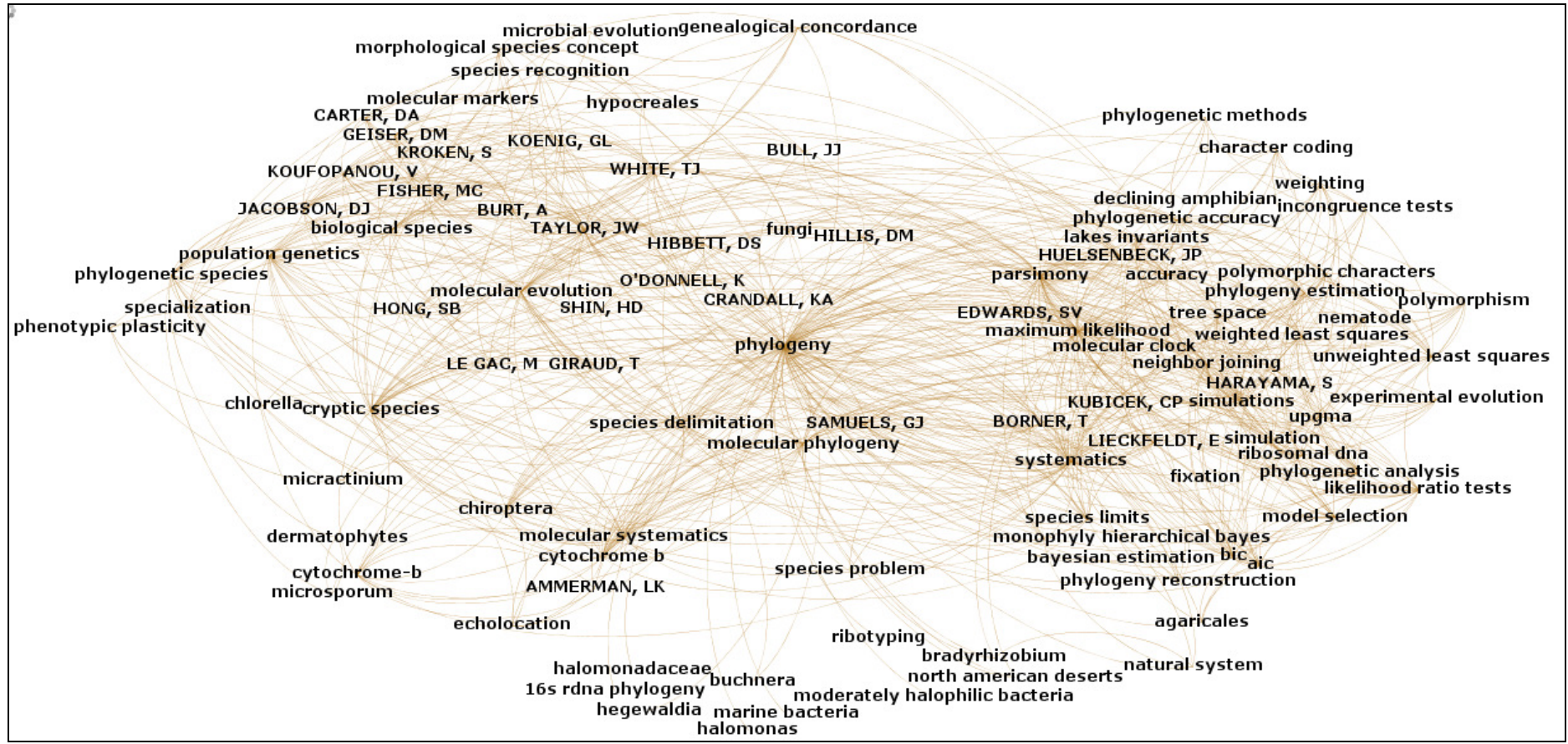

Fig 9. The recombination subgraph

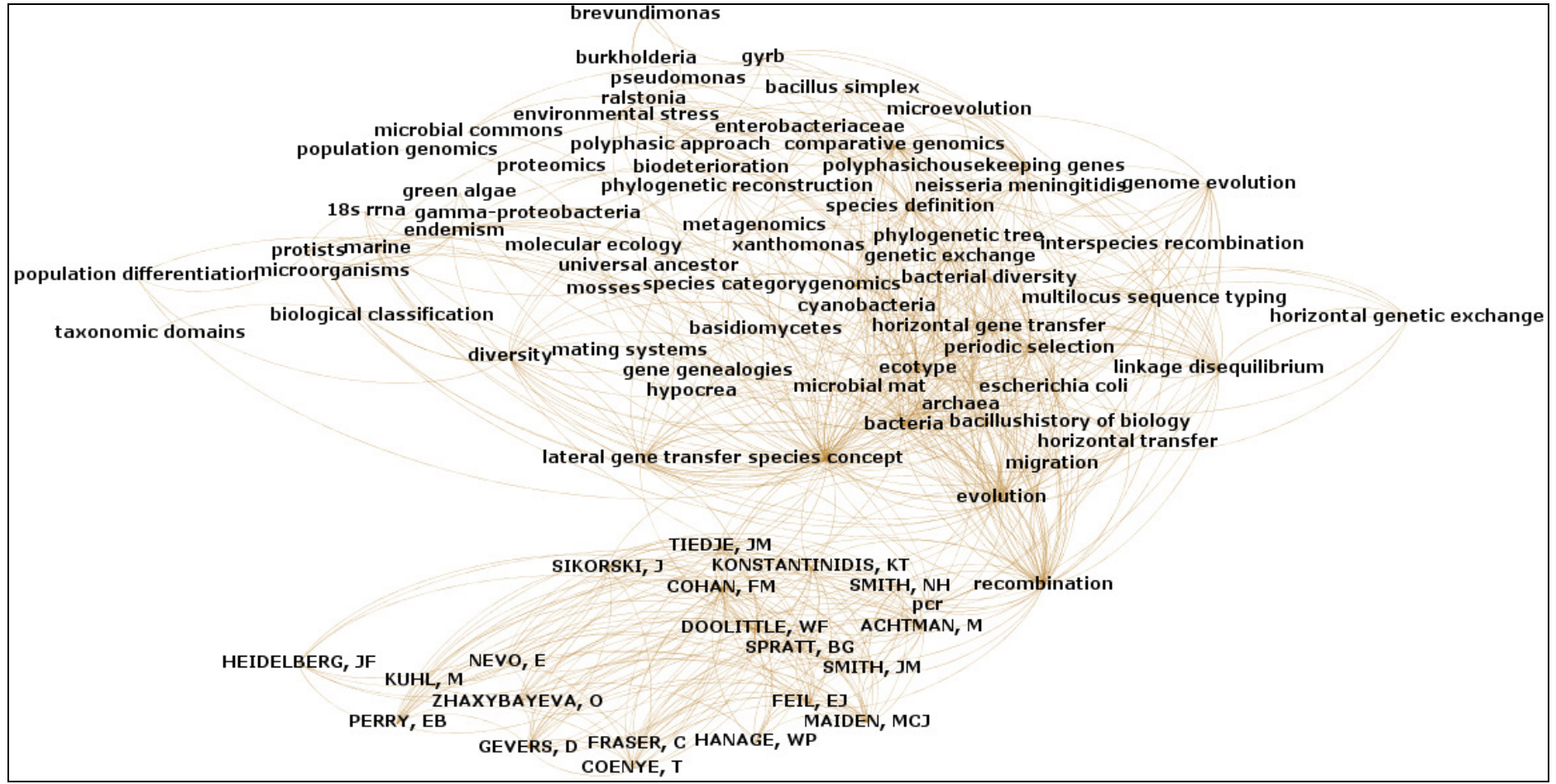

Fig 10. The speciation/BSC subgraph

Fig 11. The diatoms subgraph

**Acknowledgement**

This paper was supported by the János Bolyai Research Scholarship of the Hungarian Academy of Sciences; the European Union and the European Social Fund through project FuturICT.hu (grant no.: TÁMOP-4.2.2.C-11/1/KONV-2012-0013).